# Dynamics of Cu2O Rydberg excitons -Kerr-effect and quantum beats

**Gerard Czajkowski**

Technical University of Bydgoszcz Al. Prof. S. Kaliskiego 7 85-789 Bydgoszcz, Poland

 E-mail: czajk@pbs.edu.pl

**Abstract**. We investigate the nonlinear refraction and the nonlinear Kerr phase shift which are due to Rydberg excitons induced in $Cu_2O$ crystal by short-time pulses. We observe the phenomenon of quantum beats, analogous to that observed in exciton emission spectra, obtained in the same conditions. The calculated temporal evolution show dependence on the exciton state number, the applied laser power, and quantities related to dissipative processes.

## 1. Introduction

Since the first observation of Rydberg excitons (REs) in cuprous oxide ($Cu_2O$) [1], the physics of REs come a long way, ranging from linear optical properties in bulk crystals, to strong Kerr nonlinearities[2] (for a review see [3]). When the first period of investigation was concentrated on effects resulting from stationary excitation, recently, the dynamics of excitons has attracted large attention. In this paper we present the results of the time dependent behavior of nonlinear refraction index and Kerr phase shift, focusing the attention on the relation between the mentioned phenomena and quantum beats. Our calculation are based on the so-called real density matrix approach (RDMA), adapted to the case of short-pulse excitations, and supplemented by the time dependent functions, resulting from interference between neighboring exciton states. In Section 2 we recall the basic RDMA equations, determining the linear and nonlinear susceptibility. We also present the definitions of quantities which will be described in the following. In Section 3 we transform the presented equations to the form, which will be then applied for calculation of time dependent nonlinear optical functions. Section 4 is devoted to analysis of dissipative processes and calculation of quantities related to them. In Sections 5 and 6 we calculate the nonlinear refraction index and the Kerr phase shift in the given exciton state, showing the dependence on time, the state number, and the applied laser power. The results of calculations are presented in Section 7. We present our conclusions in Section 8. The Appendix contains supplementary information.

## 2. Theory

Recently experiments with $Cu_2O$ crystals were performed, when two waves with energies close to $E_g/2$ and finite pulse duration, excited a population of excitons in various states $E_n$ [4], **n** stands for the set of quantum numbers *n,l,m*. In what follows we consider only the S states, labelled by the principal quantum number n. The population decays then with a lifetime $\tau_n$. The emission line shapes show deviations from the decay line, which are due to interference effects between neighboring exciton states and are termed quantum beats [4, 5]. The lifetimes can be read off from the shape of time-dependent susceptibility, which consists of linear susceptibility $\chi^{(1)}$, and nonlinear ones $\chi^{(3)}$ [2, 6]

$$\chi^{(1)}(\omega,t) = (2/\varepsilon_0) \sum_n \psi_{osc,n}(t) \, (G_n(t,u)F(u))$$

$$\times \, (|c_n|^2 \, (E_n + E_g)) \, \{(E_g + E_n)^2 - \hbar^2 \, [(\omega_1 + \omega_2) + i/\tau_n]^2 \}^{-1} \, ,$$

$$\chi^{(3)}(\omega,t) = - \, (2M_0^2/\varepsilon_0) \sum_n c_n(A_n + B_{\boldsymbol{n}})(E_{\boldsymbol{n}} + E_g) \, \{[E_g + E_{\boldsymbol{n}} - (\hbar\omega_1 + \hbar\omega_2)]^2 + (\hbar/\tau_{\boldsymbol{n}})^2 \}^{-1}$$

$$\times \sum_s (T_1/\tau_s) \, c_s \, \phi_s(0) \, \{(E_g + E_s)^2 - \hbar^2 \, (\omega_1 + \omega_2 + i/\tau_s)^2 \}^{-1}$$

$$\times (G_n(t,u)F(u)) \, (G_s(u,y)F(y)) \, \psi^2_{osc,n}(t).$$

(1)

In the above equation $M_0$ is the interband-transition integrated dipole strengths, $\varepsilon_0$ the vacuum electric susceptibility constant, $\omega_1, \omega_2$ the frequencies of the impinging waves, $\phi_n$ are the excitonic eigen-functions, $E_n$ the related eigenvalues, $\tau_n$ their lifetimes, $|c_n|^2$ the corresponding oscillator strengths, and $G_n(t,u)$ the Green functions related to the exciton decline process. The quantity $T_1$ denotes the inter-band recombination time, and $A_n$, $B_n$ represent the intra-band dissipation [7]. The function $F(t)$ determines the impulse shape, and $G(t,u)$ is the Green function appropriate for the exciton decay process. The functions $\psi_{osc,n}$ describe the quantum beats. The exciton emission reported in [4] is described by the imaginary part of $\chi^{(3)}$. Having $\chi^{(1)}$ and $\chi^{(3)}$, we are also able to determine the nonlinear index of refraction $n_2$, defined as [2]

$$n_2 = \mathrm{Re} \, \chi^{(3)} (c \, \varepsilon_0 \, n_0^2)^{-1}, \qquad\qquad n_0^2 = \varepsilon_b \, (1 + \varepsilon_b^{-1} \, \mathrm{Re} \, \chi^{(1)}) \, , \qquad\qquad (2)$$

where c the vacuum light velocity, and $\varepsilon_b$ the bulk crystal dielectric constant. The total index of refraction is

$$n_{tot}^2 = \varepsilon_b + \chi^{(1)} + \chi^{(3)} |E_{prop}|^2 \, , \quad n_{tot}(I) = n_0 + n_2 I,$$

$$n_0^2 = \varepsilon_b + \chi^{(1)}, \qquad\qquad I = (c/2) \, n_0 \, \varepsilon_0 \, |E_{prop}|^2 \, . \qquad\qquad (3)$$

$E_{prop}$ being the amplitude of the electromagnetic wave. This amplitude is related to the intensity $I$ of the incoming wave by the relation

$$|E_{prop}|^2 = 2\zeta \, I \qquad\qquad\qquad (4)$$

where $\zeta \approx 377 \, \Omega$ is the impedance of free space. Note that we have to distinguish the refraction index ň from the state number n. The phase shift is calculate

$$\Delta\varphi = (\omega L/c) \, [n_{tot}(I) - n_{tot}(0)], \qquad\qquad\qquad (5)$$

where $n_{tot}(I)$ is the total refraction index obtained for average intensity $I$ inside the crystal, and L is the crystal length.

Below we discuss the power dependence dependence $n_{tot}(P)$ which is crucial for the Kerr effect (5). Both susceptibilities, linear and nonlinear, depend on P through the dependencies $\tau_n(P)$ and $A_n(P)$, $B_n(P)$.

As we will show in the following, the nonlinear optical effects depend, in the case of short-pulse excitation, on time t, and applied laser power P. The time t = 0 corresponds to the maximum of the incoming pulse. We find it opportune to use in the formulas the nominal laser power P. The dependence $\tau_n(P)$ of the exciton life time was observed experimentally [4], and can be described by the formulas [6]

$$\tau_n = \tau_0 \exp( nb_1 - n^2 b_2 - P b_3 ) \quad . \tag{6}$$

The coefficients $\tau_0, b_1, b_2, b_3$ ($\tau_0$ in ps, $b_3$ in 1/mW)

$$b_1 = 0.98, \qquad b_2 = 0.04, \qquad b_3 = 6.6 \times 10^{-3} , \qquad \tau_0 = 0.197, \tag{7}$$

were obtained from fitting experimental results from [4]. The relation $\tau_n(P)$ is illustrated in Figures 1 and 2. Since we aim to connect the Kerr effect with quantum beats, we refer to experiments [4], and consider the two-photon absorption resonant with various Rydberg states, focusing the attention on the resulting one-photon emission at twice the pump energy

$$\hbar\omega_1 + \hbar\omega_2 = E_g . \tag{8}$$

In what follows we consider the lowest even states of the so-called yellow series excitons in $Cu_2O$. For these states

$$Eg \approx 2 \text{ eV}, \qquad E_n \propto 10^{-2} \text{ eV}, \qquad 0.1 \leq \Gamma_n \leq 1 \text{ meV}, \tag{9}$$

and, in consequence, the following approximations can be made

$$\text{Re}\chi^{(1)}(\omega,t) \approx (2/\varepsilon_0) \sum_n \psi_{osc,n}(t) \ (G_n(t,u) F(u)) |c_n|^2 / 2E_n , \tag{10}$$

$$\text{Re} \ \chi^{(3)}(\omega,t) \approx - (2M_0^2/\varepsilon_0) \sum_{nm} c_n c_m (A_n + B_n) \ \phi_m(0) \ T_1 (2E_n E_m^2 \tau_m)^{-1}$$
$$\times (G_n(t,u) F(u)) \ (G_m(u,y) \ F(y)) \ \psi_{osc,n}^2(t). \tag{11}$$

In Refs. [4, 6] the effect of the pump power on the dynamics and the exciton lifetime was studied for each state (separately). We perform the calculations in analogous situation, therefore we drop the summation in Eqs. (1,12) obtaining

$$\text{Re}\chi^{(1)}(\omega;t) \approx (2/\varepsilon_0) \ \psi_{osc,n}(t) \ (G_n(t,u) F(u)) |c_n|^2 / 2E_n ,$$

$$\text{Re} \ \chi^{(3)}(\omega,t) \approx - (2M_0^2/\varepsilon_0) |c_n|^2 (A_n + B_n) \ \phi_n(0) \ T_1 T_1 (2E_n E_m^2)^{-1}$$
$$\times (G_n(t,u) F(u)) \ (G_m(u,y) \ F(y)) \psi_{osc,n}^2(t). \tag{12}$$

For the further calculations we must define all the terms arriving in the above equations. We recall the definitions following Ref. [6]. The coefficients $c_{nlm}$ and eigenfunctions $\phi_{nlm}$ are defined as follows

$$c_{nlm} = \int d^3r \ \mathbf{M}(\mathbf{r}) \ \phi_{nlm}(\mathbf{r}), \qquad \phi_{nlm} = R_{nl}(r) \ Y_{lm}(\theta,\varphi), \tag{13}$$

$Y_{lm}(\theta,\varphi)$ are the spherical harmonics, $R_{nl}(r)$ are the radial functions of a corresponding hydrogen atom-like Schrödinger equation. The function $\mathbf{M}(\mathbf{r})$ is the so-called transition dipole density [8], and describes the quantum coherence between the macroscopic electromagnetic field (the wave propagating in the crystal), and the inter-band transitions. For the S type transitions it has the form

$$\mathbf{M(r)} = \mathbf{M_0} \, Y_{00}^2 (1/rr_0^2) \exp(-r/r_0), \tag{14}$$

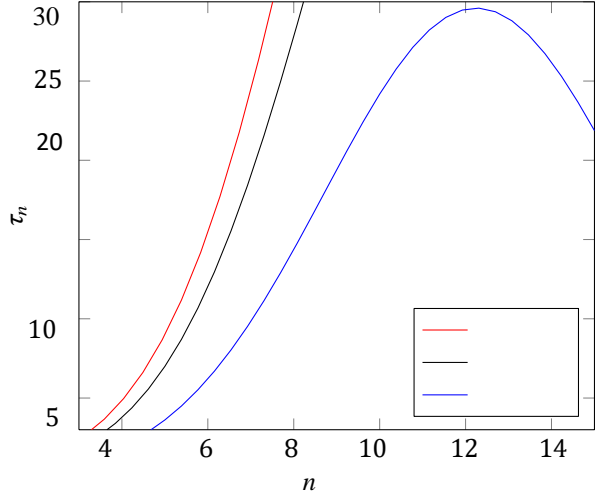

**Figure 1.** Exciton life-time from Eq. (6), for three values of the applied power, 12 mW, (red line), 50 mW, (black line) and 150 mW. (blue line)

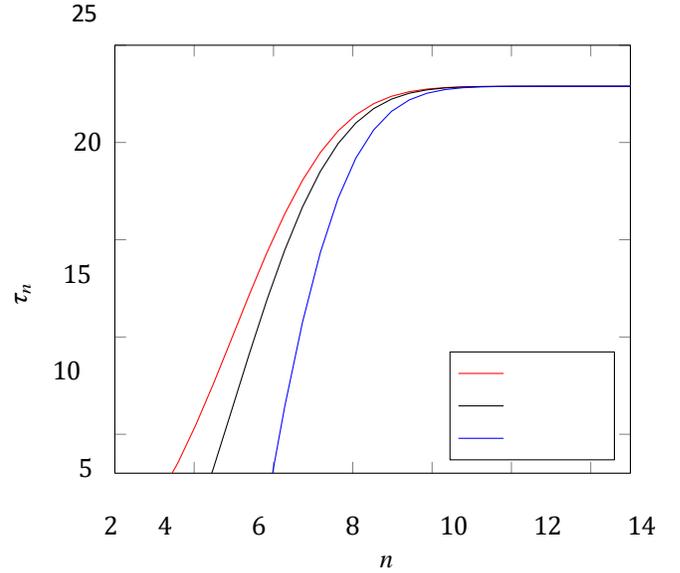

**Figure2.** Exciton life-time from Eq. (8) and showing the effect of saturation, the applied powers as in **Fig**. 1.

where $r_0$ is the so-called coherence radius, defined as $r_0 = (2\mu E_g/\hbar^2)^{-1/2}$, and $\mu$ being the reduced electron-hole effective mass. The exciton oscillator strength reads

$$f_n = n(n - r_0/a^*)^{2(n-1)} \, (n + r_0/a^*)^{-2(n+1)}, \tag{17}$$

where $a^*$ is the exciton effective Bohr radius. As can be seen from Figure 3, for large n the approximation

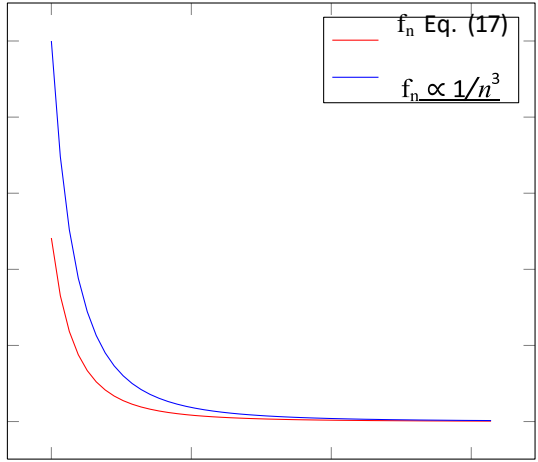

**Figure 3**. Oscillator strength

$f_n \approx 1/n^3$ can be used (red curve calculated with $r_0 = 0.2\ a^*$). From the formulas (14-16) we get

$$|c_n|^2 = (M_0^2/\pi)\ f_n, \qquad (19)$$

where

$$M_0^2 = \varepsilon_0\ \varepsilon_b \pi\ a^{*3}\ \Delta_{LT}/2f_{100}\ , \qquad (20)$$

with the longitudinal-transversal energy $\Delta_{LT}$.

The above calculated quantities, inserted into equations (1,12) give the relations

$$\mathrm{Re}\ \chi_n^{(1)}(\omega,t) = \varepsilon_b\ (\Delta_{LT}/2E_n)\ (f_n/f_1)\ ((G_n(t,u)F(u))\psi^2_{\mathrm{osc},n}(t),$$

$$\mathrm{Re}\ \chi_n^{(3)}(\omega,t) = -(( \Delta_{LT}/2E_n)\ (f_n/f_1)\ \chi_{0n}^{(3)}\ (G_n(t,u)F(u))(G_n(u,y)F(y))\psi^2_{\mathrm{osc},n}(t),$$

$$\chi_{0n}^{(3)} = \varepsilon_0\ \varepsilon_b \pi\ \Delta_{LT}\ [a^{*3}(A_n + B_n)\ \phi_n(0)\ T_1]\ (2E_n^2\tau_n)^{-1}\ . \qquad (21)$$

The quantities $A_n$, $B_n$ are related to nonlinear dissipative processes (for example [2], [7]), and are defined as

$$A_n = <\phi_n(\mathbf{r})|f_{0e}(\mathbf{r})>, \qquad B_n = <\phi_n(\mathbf{r})|f_{0h}(\mathbf{r})>, \qquad (22)$$

where $f_{0e}$, $f_{0h}$ are the normalized Boltzmann distributions for electrons and holes, respectively. The expressions (21) are then used in Eq. (2), giving the formulas for the refraction indices

$$n_{0n}^2 = \varepsilon_b\ [1 + C_n^{(1)}F_n^{(1)}(t)],$$

$$n_{2n} = C_n^{(2)}F_n^{(1)}(t)],$$

$$C_n^{(1)} = (\Delta_{LT}/E_n)\ (f_n/f_1), \qquad (23)$$

$$F_n^{(1)}(t) = ((G_n(t,u)F(u))\psi^2_{\mathrm{osc},n}(t),$$

$$C_n^{(2)} = C_n^{(1)} \Delta_{LT} a^{*3} [a^{*3}(A_n + B_n) \phi_n(0) T_1] (2cE_n^2 \tau_n)^{-1} ,$$

$$F_n^{(2)}(t) = [1 + C_n^{(1)} F_n^{(1)}(t)]^{-1} (G_n(t,t')F(t'))(G_n(t'',t')F(t'))\psi^2_{osc,n}(t),$$

with the dimension $[C_n^{(2)}] = m^2/W$. The equations (23) define the refraction indices for given exciton state n, and in short-pulse excitation regime. As can be seen, they reveal oscillatory behavior (quantum beats), described by the time-dependent terms $F_n^{(1)}(t)$, $F_n^{(2)}(t)$. The quantities $C_n^{(1)}$, $C_n^{(2)}$ define amplitude of oscillations. Having calculated the refraction indices, one can determine the phase shift (5), which also will show oscillatory behavior at the given state.

### 3. Calculations of coefficients $A_n$, $B_n$

As follows from expressions (21), almost all ingredients are known, besides of the quantities $A_n$, $B_n$. Below we consider the exciton S states, thus the definitions (22) take the form

$$A_n = \int r^2 dr \, R_{n0} \exp(-r^2/2 \, \lambda^2_{th,e}), \qquad B_n = \int r^2 dr \, R_{n0} \exp(-r^2/2\lambda^2_{th,h}), \qquad (24)$$

where $\lambda_{th,e}$, $\lambda_{th,h}$ are the so-called thermal lengths for electrons and holes, respectively [2], [7],

$$\lambda_{th,e} = (\hbar^2/m_e \, k_B T)^{1/2} , \qquad \lambda_{th,h} = (\hbar^2/m_h \, k_B T)^{1/2} , \qquad (25)$$

$k_B$ being the Boltzmann constant, and $T$ is the temperature, $k_B = 8.617 \times 10^{-2}$ meV/K. The coefficients $A_n$, $B_n$ depend on the temperature by definitions of $\lambda$'s (25), which can be written in the form

$$\underline{\lambda}_{th,e} = \lambda_{th,e}/a^* = (1/a^*)(2\mu \, \hbar^2/m_e(2\mu) \, k_B T)^{1/2} = (2 \, \mu R^*/m_e \, k_B T)^{1/2} ,$$

$$\underline{\lambda}_{th,h} = \lambda_{th,h}/a^* = (1/a^*)(2\mu\hbar^2/m_h(2\mu) \, k_B T)^{1/2} = (2 \, \mu R^*/m_h \, k_B T)^{1/2} .$$

The exciton Rydberg energy $R^*$, and the excitonic Bohr radius $a^*$, were obtained by the formulas

$$R^* = 13600 \, \mu/ \, \varepsilon_b^2, \qquad a^* = (m_0/\mu) \, \varepsilon_b \times 0.0529 \text{ nm} \qquad (26)$$

**Table 1**. Band parameter values for Cu$_2$O from [9],
Energies in meV, masses in free electron mass $m_0$,
lengths in nm, $\Delta_{LT}$ from [10].

| Parameter | Value |
|---|---|
| $E_g$ | 2172.08 |
| R* | 90.88 |
| $\Delta_{LT}$ | 5x 10-2 |
| $m_e$ | 0.985 |
| $m_h$ | 0.575 |
| $\mu$ | 0.363 |
| $\varepsilon_b$ | 7.37 |
| a* | 1.07 |

Using the parameter values from Table 1, we obtain

$$\underline{\lambda}_{th,e} = 27.88 \, (K/T)^{1/2} \qquad (2 \, \underline{\lambda}_{th,e})^{-2} = 6.43 \times 10^{-4} \, (T/K)$$

$$\underline{\lambda}_{th,h} = 36.49 \ (K/T)^{1/2} \qquad (2\underline{\lambda}_{th,h})^{-2} = 3.75 \ \times 10^{-4} \ (T/K).$$

The above quantities are then used in the determination of $A_n$, $B_n$ by equations (24). The results are given below in Tables 2-4. As it can be seen from the data displayed in the Tables, the coefficients $A_n$, $B_n$ have alternative signs. It reflects the oscillatory behavior of the Lagrange polynomials, being constituents of the radial functions $R_{nl}$.

**Table 2**. Quantities $A_4$, $A_4 \phi_4(0)$, $B_4$, $B_4 \phi_4(0)$, and $(A_4+B_4)_0 = (A_4+B_4) \phi_4(0)$, for 5 values of temperature

| Temp. | $A_4$ | $A_4 \phi_4(0)$ | $B_4$ | $B_4 \phi_4(0$ | $(A_4+B_4)_0$ |
|--------|--------|-----------------|--------|----------------|---------------|
| 4.2 | $-10.33$ | $-2.58$ | $-26.94$ | $-6.73$ | $-9.31$ |
| 10 | 1.138 | 0.288 | $-3.48$ | $-0.87$ | $-0.59$ |
| 20 | 0.905 | 0.248 | 1.5 | 0.38 | 1.38 |
| 30 | 0.20 | 0.05 | 1.3 | 0.33 | 0.38 |
| 40 | $-0.237$ | $-0.0638$ | 0.67 | 0.17 | 0.11 |

**Table 3**. Quantities $A_5$, $A_5 \phi_5(0)$, $B_5$, $B_5 \phi_5(0)$, and $(A_5+B_5)_0 = (A_5+B_5) \phi_5(0)$, for 5 values of temperature

| Temp. | $A_5$ | $A_5 \phi_5(0)$ | $B_5$ | $B_5 \phi_5(0$ | $(A_5+B_4)_0$ |
|--------|--------|-----------------|--------|----------------|---------------|
| 4.2 | $-19.36$ | $-2.88$ | $-26.4$ | $-3.933$ | $-6.81$ |
| 10 | $-5.57$ | $-0.83$ | $-13.13$ | $-1.96$ | $-2.79$ |
| 20 | $-1.3$ | $-0.194$ | $-4.06$ | $-0.6$ | $-0.794$ |
| 30 | -0.712 | -0.1 | -1.8 | -0.26 | $-0.36$ |
| 40 | $-0.54$ | $-0.0838$ | $-1$ | $-0.145$ | $-0.225$ |

**Table 4**. Quantities $A_6$, $A_6 \phi_6(0)$, $B_6$, $B_6 \phi_6(0)$, and $(A_6+B_6)_0 = (A_6+B_6) \phi_6(0)$, for 5 values of temperature

| Temp. | $A_6$ | $A_6 \phi_6(0)$ | $B_6$ | $B_6 \phi_6(0)$ | $(A_6+B_6)_0$ |
|--------|--------|-----------------|--------|-----------------|---------------|
| 4.2 | $-0.666$ | $-0.09$ | 0.469 | 0.065 | $-0.025$ |
| 10 | 0.225 | 0.03 | $-0.63$ | $-1.96$ | $-2.79$ |
| 20 | 0.48 | 0.066 | 0.423 | 0.0588 | 0.125 |
| 30 | 0.187 | 0.026 | 0.55 | 0.076 | 0.1 |
| 40 | $-0.033$ | $-0.004$ | 0.372 | 0.052 | 0.048 |

## 4. Calculation of the nonlinear refraction index

Using Eqs. (23), we calculate the time dependence of the nonlinear refraction index for S exciton states in a $Cu_2O$ crystal. In calculations we use a Gaussian shaped normalized pulse

$$F(t) = F_{max} (2\pi\tau_p^2)^{-1/2} \exp(-t^2/2 \ \tau_p^2) \qquad (28)$$

where $\tau_p$ is the pulse temporal duration. Using this shape we calculate the expressions including the Green function G,

$$G_n(t,u)=(\tau_n/2) \exp(-|t-u|/\tau_n). \qquad (29)$$

In the lowest approximation we obtain

$$(G_n(t,u)F(u)) \approx F_{max} (\tau_n/2) \exp(-|t|/\tau_n). \qquad (30)$$

The nonlinear refraction index for the state n has the form

$$n_{2n}(t) \; I = X_{0n}^{(3)} \; (G_n(t,u)F(u))(G_n(u,y)F(y)) \; \beta^2_{osc,n} (t), \qquad (31)$$

where

$$X_{nP}^{(3)} = - 0.795 \times 10^3 \, P(mW) \, (\Delta_{LT} f_n / \, E_n \, f_1)$$

$$\times \; (\Delta_{LT} a^{*3}/2cE_n^2)[ \; (A_n + B_n)_0] \; \Psi_n^2 (10^3 \, t_1/\tau_n) \qquad (32)$$

$$= -n^3 \, (4.25x \; 10^{-17}) \; \Psi_n^2][(A_n + B_n)_0] \, t_1/\tau_n) \, \exp(b_3P) \quad [mm^2 /mW],$$

with $t_1 = 10^{-3} \, T_1$, $\beta_{osc,n}$ and $\Psi_n$ will be defined below (see (A1)). The matrix elements $X_{nP}^{(3)}$ are displayed in Table 5. The total refraction index will be calculated by Eqs. (3). First we calculate the index $n_0$

$$n_{0n} \approx \sqrt{\varepsilon_b + \chi_{0n}^{(1)}} \; \psi_{osc,n} \, (t) \; (G_n(t,u)F(u)),$$

$$\chi_{0n}^{(1)} = \Delta_{LT} \Psi_n f_n \, (2 \, E_n \sqrt{\varepsilon_b f_1})^{-1} = 10^{-4} \, (1/n) \, \Psi_n \, . \qquad (33)$$

When using Eqs. (3), one must to insert the laser power per unit area. Having in mind the experiments in Ref. [4], where the laser with 20 μm spot diameter is focused on the sample, we obtain

$$mW/(\pi(20 \; \mu m)^2) = (109/ \, 400 \; \pi)(W/m^2) = 0.795 \times 10^6)(W/m^2),$$

$$I = P \, (mW) \times 0.795 \times 10^3 \qquad (mW/mm^2). \qquad (34)$$

## 6. Calculation of the phase shift

Using Eqs. (3) and (5) we calculate

$$n_{tot,n}(I) = n_{0n} + n_{2n}I, \qquad (35)$$

$$n_{tot,n}(0) = \sqrt{\varepsilon_b + \chi_{0n}^{(1)}} \; \psi_{osc,n} \, (t) \, \exp(-|t|/\tau_{n, (P=0)}).$$

The above result is now used to calculate $n_{tot,n}(I)$-$n_{tot,n}(P=0)$

$$n_{tot,n}(I) - n_{tot,n}(P=0) = \Delta \, n_{tot,n} + n_{2n} \, I, \qquad (36)$$

$$\Delta \, n_{tot,n} = [ \exp(-|t|/\tau_{n, P}) - \exp(-|t|/\tau_{n, (P=0)}) \; ] \; \psi_{osc,n} \, (t).$$

Thus the final expression for the phase shift in the exciton state $n$ has the form

$$\Delta\phi_{n,P} = (\omega L/c)[n_{tot,n}(I) - n_{tot,n}(P=0)] = (\omega L/c)(\Delta n_{tot,n} + n_{2n} I) \quad. \tag{37}$$

## 7. Results

Using the expressions (35,37), with definitions (A.1) – (A.3), we obtain the formulas for the time dependence of the refraction indices $n_2$, $n(P)$, and the phase shifts $\Delta\phi_4 - \Delta\phi_6$. The line-shapes of the nonlinear refraction indices, for three applied laser powers, are presented in Figs. 6-8. We observe the general tendency of increasing the value with increasing state number and applied power. Simultaneously, the quantum beats appear. Their amplitudes decrease with time and the beats disappear after about 20 ps. In two cases we observe the inversion of sign of the indices with the increasing power (the states $n = 4,6$)

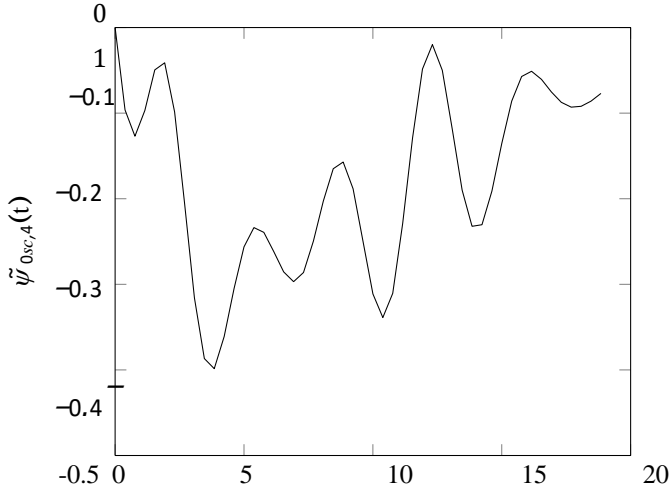

**Figure 4**. Function $\tilde{\psi}_{osc;4} \times 9.6 \times [\exp(-t/4.83) - \exp(-t/5.23)]$

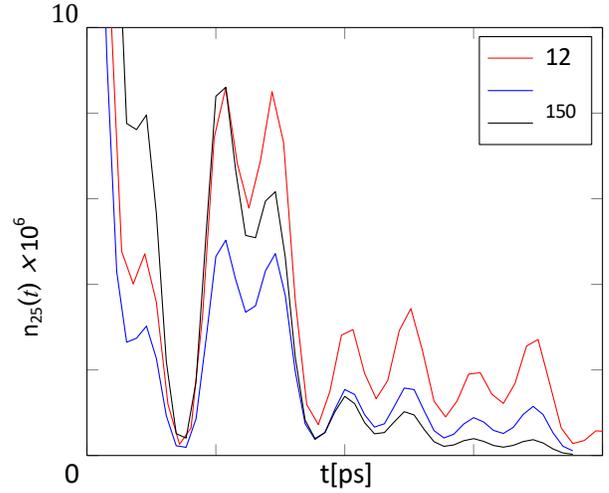

**Figure 6**. Nonlinear refraction index $n_{2,5,12}$ (red), $n_{2,5,50}$ (blue), $n_{2,5,150}$ (black)

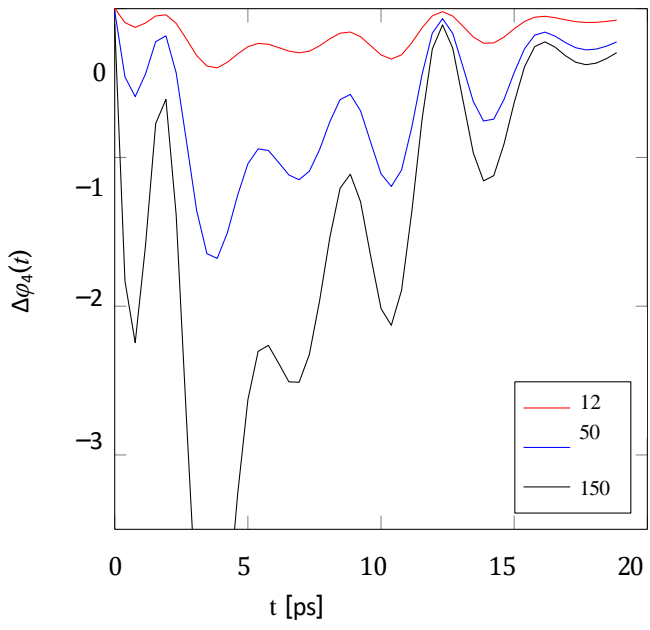

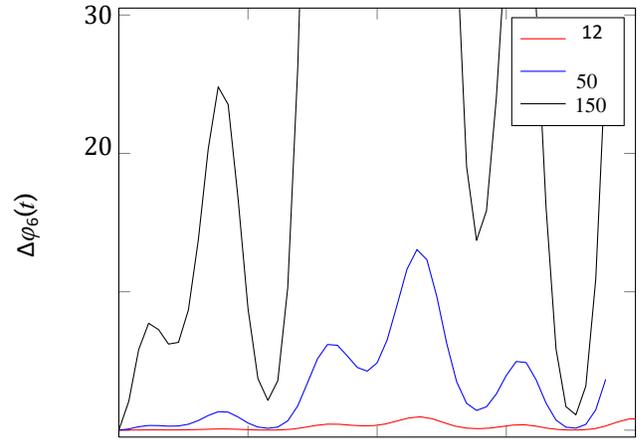

**Figure 8.** Phase shifts $\Delta\varphi_{4,12}$, $\Delta\varphi_{4,50}$, $\Delta\varphi_{4,150}$

**Figure 10.** Phase shifts $\Delta\varphi_{6,12}$, $\Delta\varphi_{6,50}$, $\Delta\varphi_{6,150}$

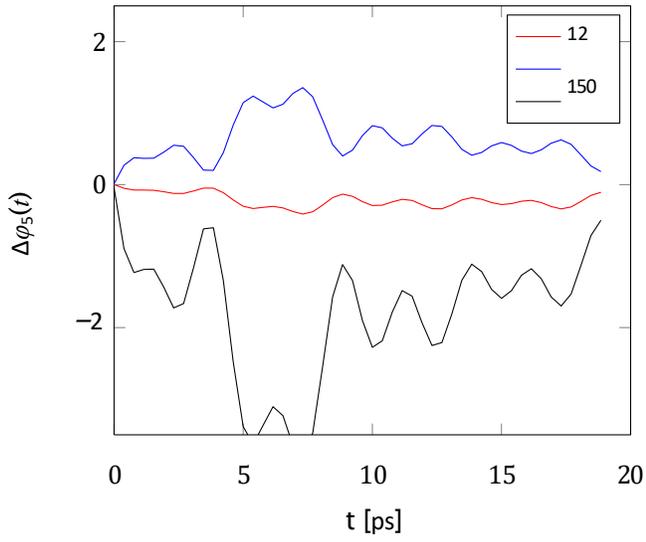

**Figure 9.** Phase shifts $\Delta\varphi_{5,12}$, $\Delta\varphi_{5,50}$, $\Delta\varphi_{5,150}$

## 8. Conclusions

We presented a method based on the so-called real density matrix approach, to describe the time evolution of the nonlinear refraction index and the Kerr phase-shift, induced in $Cu_2O$ crystal by short-time pulses. We analyzed the case of two-photon excitation of S excitons, for various excitation power. The calculated line shapes show quantum beats, where the frequencies are the same as those observed in emission spectra. It is due to the fact, that the frequencies depend only on involved excitonic eigenenergies. The amplitudes of the oscillations show a quite complicated dependence on the exciton state number and the applied laser power. The Kerr-effect for $Cu_2O$ S excitons was investigated previously in the case of stationary excitation. To our best knowledge, the time evolution was not investigated. Therefore, a comparison of our theoretical results with experiment is lacking. Future experimental results, together with the presented theory, can give an insight into the role of dissipative processes, which, to a considerable extent, determine the quantum beats shape.

## Appendix A. Supplemental information

## Appendix A.1. Oscillatory functions $\psi_{osc,n}$

The oscillatory functions for the exciton states n = 4,5,6 $\psi_{osc,n}$, used through calculations, have the form [6]

$$\psi_{osc,4} = \Psi_4 \times \beta_{osc,4}(t),$$

$$\beta_{osc,4}(t) = 1 + 0.32 \cos(1.3t) + 0.4 \cos(1.76t) + 0.1 \cos(2.29t) + 0.132 \cos(1.85t),$$

$$\Psi_4 = 700, \qquad\qquad (A.1)$$

$$\psi_{osc,5}(t) = \Psi_5 \times \beta_{osc,5}(t),$$

$$\beta_{osc,5}(t) = 1 + 0.285 \cos(0.829t) + 0.356 \cos(1.068t) + 0.277 \cos(2.51t) + 0.036 \cos(2.29t)$$

$$+ 0.045 \cos(1.85t), \qquad\qquad \Psi_5 = 713, \qquad (A.2)$$

$$\psi_{osc,6}(t) = \Psi_6 \, \beta_{osc,6}(t),$$

$$\beta_{osc,6}(t) = 1 + 0.49 \cos(0.58t) + 0.37 \cos(1.6t) + 0.077 \cos(1.2t) + 0.09 \cos(0.845t),$$

$$\Psi_6 = 2211. \qquad\qquad (A.3)$$

The arguments of trigonometric functions are related to the exciton beatings characteristic frequencies given in Tables 6 and 7.

**Table 5**. Matrix elements $X_{nP}^{(3)}$ in units $10^{-6}$.

| n/P | 12 | 50 | 150 |
|-----|------|-------|------|
| 4 | 1.55 | -8.31 | 48 |
| 5 | 8.0 | 5.52 | 20 |
| 6 | 1.39 | -14.8 | -40 |

**Table 6.** Exciton beatings characteristic frequencies $\Omega_{n1n2}$, corresponding to $(n_{2D} \rightarrow n_{1S})$ transitions.

| $n_{2D}/n_{1S} \rightarrow$ $\downarrow$ | 3 | 4 | 5 | 6 | 7 |
|------|------|-------|-------|------|------|
| 3,1 | 2.3 | 4 | | | |
| 3,2 | 3.06 | | | | |
| 4,1 | | 1.3 | 2.29 | | |
| 4,2 | | 1.745 | 1.851 | | |
| 5,1 | | | 0.835 | 1.12 | |
| 5,2 | | | 1.11 | 0.845 | |
| 6 | | | 2.53 | 0.58 | 0.59 |
| 7 | | | | 1.6 | 0.42 |

**Table 7.** Exciton beatings characteristic frequencies $\Omega_{n1n2}$, corresponding to $(n_{2D} \rightarrow n_{1S})$ transitions, for $n_{1,2} = 9\text{-}11,13\text{-}15$

| $n_{2D}/n_{1S} \rightarrow$ $\downarrow$ | 9 | 10 | 11 | 13 | 14 | 15 |
|------|-------|-------|-------|-------|-------|-------|
| 9 | 0.346 | 0.167 | 0.32 | | | |
| 10 | 0.7 | 0.15 | 0 | | | |
| 11 | 0.926 | 0.395 | 0.243 | | | |
| 13 | | | | 0.172 | 0.035 | 0.07 |
| 14 | | | | 0.29 | 0.15 | 0.046 |
| 15 | | | | 0.386 | 0.24 | 0.14 |

## Appendix A.2. Kerr phase shifts, formulas

$$\Delta\varphi_{4,12} \tag{A.4}$$

$\Delta\varphi_{4,12}(t) = 550\{\ 0.0175[\ \exp(-|t|/4.83) - \exp(-|t|/5.23)]$
$\times[1+0.32\cos(1.3t)+0.4\cos(1.76t)+0.1\cos(2.29t)+0.132\cos(1.85t)]$
$+1.55\times10^{-6}\exp(-|t|/4.83)$
$\times[\ 1+0.32\cos(1.3t)+0.4\cos(1.76t)+0.1\cos(2.29t)+0.132\cos(1.85t)]^2\ \},$

where we the notation '$\Delta\varphi_{4,12}$' means: the Kerr phase shift for the exciton in state n=4, and the applied laser power equal 12 mW,

$$\Delta\varphi_{4,50} \tag{A.5}$$

$\Delta\varphi_{4,50}(t) = 550\{0.0175[\exp(-|t|/3.76) - \exp(-|t|/5.23)]$
$x\ [1+0.32\cos(1.3t)+0.4\cos(1.76t)$
$+0.1\cos(2.29t)+0.132\cos(1.85t)] -8.31\times10^{-6}\exp(-|t|/3.76)\times[\ 1+0.32$
$\cos(1.3t)+0.4\cos(1.76t)+0.1\cos(2.29t)+0.132\cos(1.85t)]^2\ \},$

$$\Delta\varphi_{4,150} \tag{A.6}$$

$\Delta\varphi_{4,150}(t) = 550\{0.0175[\exp(-|t|/1.94) - \exp(-|t|/5.23)]$
$x\ [1+0.32\cos(1.3t)+0.4\cos(1.76t)\qquad +0.1\cos(2.29t)$
$+0.132\cos(1.85t)] -4.8\times10^{-5}\exp(-|t|/1.94)\times[\ 1+0.32\cos(1.3t)+0.4$
$\cos(1.76t)+0.1\cos(2.29t)+0.132\cos(1.85t)]^2\ \},$

$$\Delta\varphi_{5,12} \tag{A.7}$$

$\Delta\varphi_{5,12}(t) = 550\{0.01426[\exp(-|t|/9) - \exp(-|t|/9.73)]$
$x\ [1+0.285\cos(0.829t)+0.356\cos(1.068t)$
$+0.277\cos(2.5t)+0.036\cos(2.29t)+0.045\cos(1.85t)]$
$+4.0\times10^{-6}\exp(-|t|/9)[1+0.285\cos(0.829t)+0.356\cos(1.068t)$
$+0.1+0.277\cos(2.5t)+0.036\cos(2.29t)+0.045\cos(1.85t)]^2\},$

$$\Delta\varphi_{5,50} \tag{A.8}$$

$\Delta\varphi_{5,50}(t) = 550\{0.01426[\exp(-|t|/7) - \exp(-|t|/9.73)]$
$x\ [1+0.285\cos(0.829t)+0.356\cos(1.068t)$
$+0.277\cos(2.5t)+0.036\cos(2.29t)+0.045\cos(1.85t)]$
$+2.76\times10^{-6}\exp(-|t|/7)[1+0.285\cos(0.829t)+0.356\cos(1.068t)$
$+0.1+0.277\cos(2.5t)+0.036\cos(2.29t)+0.045\cos(1.85t)]^2\},$

$$\Delta\varphi_{5,150} \tag{A.9}$$

$\Delta\varphi_{5,150}(t) = 550\{0.01426[\exp(-|t|/3.61) - \exp(-|t|/9.73)]$
x $[1 + 0.285\cos(0.829t) + 0.356\cos(1.068t)$
$+ 0.277\cos(2.5t) + 0.036\cos(2.29t) + 0.045\cos(1.85t)]$
$+ 10^{-5}\ \exp(-|t|/3.61)[1 + 0.285\cos(0.829t) + 0.356\cos(1.068t)$
$+ 0.1 + 0.277\cos(2.5t) + 0.036\cos(2.29t) + 0.045\cos(1.85t)]^2\}$,

$$\Delta\varphi_{6,12} \tag{A.10}$$

$\Delta\varphi_{6,12}(t) = 550\{0.03685[\exp(-|t|/15.43) - \exp(-|t|/16.7)]$
x $[1 + 0.49\cos(0.58\,t) + 0.37\cos(1.6\,t) + 0.077\cos(1.2t)$
$+ 0.09\cos(0.845\,t)]$
$+ 0.69\,\text{x}10^{-6}\ \exp(-|t|/15.43)[1 + 0.49\cos(0.58\,t) + 0.37\cos(1.6\,t)$
$+ 0.077\cos(1.2t) + 0.09\cos(0.845\,t)]^2\}$,

$$\Delta\varphi_{6,50} \tag{A.11}$$

$\Delta\varphi_{6,50}(t) = 550\{0.03685[\exp(-|t|/12) - \exp(-|t|/16.7)]$
x $[1 + 0.49\cos(0.58\,t) + 0.37\cos(1.6\,t) + 0.077\cos(1.2t)$
$+ 0.09\cos(0.845\,t)]$
$- 10^{-7}\exp(-|t|/12)$ x $[1 + 0.49\cos(0.58\,t) + 0.37\cos(1.6\,t)$
$+ 0.077\cos(1.2t) + 0.09\cos(0.845\,t)]^2\}$,

$$\Delta\varphi_{6,150} \tag{A.12}$$

$\Delta\varphi_{6,150}(t) = 550\{0.03685[\exp(-|t|/6.2) - \exp(-|t|/16.7)]$
x $[1 + 0.49\cos(0.58\,t) + 0.37\cos(1.6\,t) + 0.077\cos(1.2t)$
$+ 0.09\cos(0.845\,t)]$
$- 0.048\text{x}10^{-6}\exp(-|t|/6.2)$ x $[1 + 0.49\cos(0.58\,t) + 0.37\cos(1.6\,t)$
$+ 0.077\cos(1.2t) + 0.09\cos(0.845\,t)]^2\}$.